\documentclass[conference]{IEEEtran}
\IEEEoverridecommandlockouts
\usepackage{cite}
\usepackage{amsmath,amssymb,amsfonts}
\usepackage{algorithmic}
\usepackage{graphicx}
\usepackage{textcomp}
\usepackage{xcolor}
\usepackage{caption}
\usepackage[squaren,Gray]{SIunits}
\usepackage[linesnumbered,ruled,vlined]{algorithm2e}
\usepackage{comment}

\def\BibTeX{{\rm B\kern-.05em{\sc i\kern-.025em b}\kern-.08em
    T\kern-.1667em\lower.7ex\hbox{E}\kern-.125emX}}

\SetCommentSty{mycommfont}

\SetKwInput{KwInput}{Input}                
\SetKwInput{KwOutput}{Output}              

\begin{document}

\title{Systematic high-level design of a fifth order                       Continuous-Time CRFF Delta Sigma ADC\\
\thanks{ \textbf{978-1-7281-7670-3/21/\$31.00 ©2021
IEEE}}
}

\author{\IEEEauthorblockN{1\textsuperscript{st} Abderrahmane GHIMOUZ, 2\textsuperscript{nd} Fatah RARBI, 3\textsuperscript{rd} Olivier ROSSETTO}
\IEEEauthorblockA{\textit{Univ. Grenoble Alpes, Grenoble INP, CNRS, LPSC-IN2P3} \\
38000 Grenoble, France\\
ghimouz@lpsc.in2p3.fr}
}

\maketitle

\begin{abstract}
In this paper we present the development of a Systematic high level design model based on MATLAB scripts. It is integrated into a graphical behavioral model toolbox for the synthesis and simulation of a Continuous-Time Delta Sigma ADC. For this, we decided to use a Model-based design approach which it is adopted to address problems associated with designing complex control and signal processing systems such as the case of Continuous-Time Delta Sigma ADC. The goal of our study is the design of a 10 bit ENOB ADC for energy measurement systems used in particle identification through a new generation of detectors based on diamond. Results of the synthesis of a proposed fifth order Continuous-Time Delta Sigma ADC modulator for 10-bit ENOB ADC based on a Cascaded Resonators Feedforward architecture and simulations of the dispersion of its components (due to fabrication process) using the proposed tool are demonstrated and discussed.
\end{abstract}

\begin{IEEEkeywords}
Analog to digital converters, Continuous-time, Delta Sigma, CMOS analog integrated circuits, Mathematical modeling,  front-end electronics, model-based design, circuit simulation. 
\end{IEEEkeywords}

\section{Introduction}
\begin{figure}[htbp]
\centerline{\includegraphics[scale=0.6]{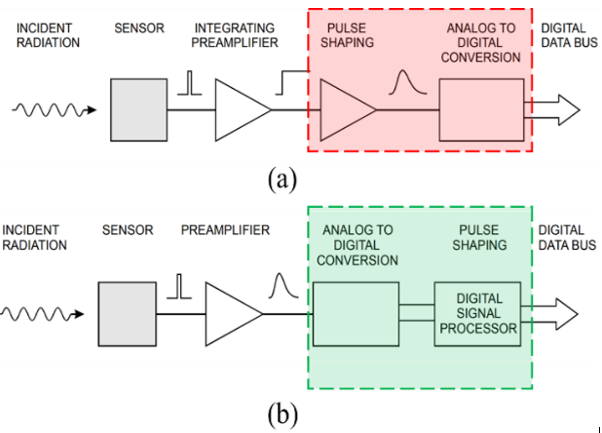}}
\captionsetup{singlelinecheck=off}
\caption[a b]{ 
 \begin{itemize}
    \item (a) Simplified diagram of the classical systems
    \item (b) Simplified diagram of the novel systems
  \end{itemize}}
\label{fig_01}
\end{figure}

Designing complex Continuous-Time (CT) Delta sigma ($\Delta \Sigma $) ADC involves engineers who possess the knowledge, intuition, creativity, and technical expertise to turn their ideas into real-life devices \cite{b1}. Until now, this type of architectures does not appear in the field of designing electronics for particle detection. Yet, we believe that it can provide interesting solution, as we will be showing in the case of energy measurement systems. \\
This class of systems are classically designed as shown in figure \ref{fig_01} (a). The problem with this topology is the analog shaping block, which represents a bottleneck when it comes to ensuring high accuracy and minimal variation between channels. It also limits the single chip integration with complexity, high power consumption and large area occupation. The idea here (for the second topology) is to digitize the signal as soon as possible as demonstrated in figure \ref{fig_01} (b). To enable that, $ \Delta  \Sigma $ ADC are recommended as shown in \cite{b2,b3}.
It was explained in the literature that these architectures – more precisely continuous time – are suitable for this kind of applications \cite{b4, b5}. The main difficulty remains in the complexity of the design of such architecture, since it requires frequency domain analysis that usually requires a lot of time, experience and effort. For this purpose, we propose in this article a systematic high-level design of a fifth order CT $ \Delta  \Sigma $ modulator for 10-bit ENOB ADC based on a Cascaded Resonators Feedforward (CRFF) architecture. To synthesis and design efficiently this modulator, we decided to adopt a model-based design approach built around the use of graphical behavioral models \cite{b1} that integrate the widely known Schreier’s Toolbox \cite{b6}, then adding a set of custom scripts. This method allows firstly the synthesis of a CT $ \Delta  \Sigma $ modulator enabling different choices of specifications, then to simulate - given a set of conditions - the effect of the variations of each main parameter of the chosen topology. This permits us to extract information about the critical parameters so we can optimize the design of each constructive block. All this development is based on MATLAB scripts applied to the graphical model and the conditions of simulations in SIMULINK. \\ 
For this paper, we will focus on the study the effect of component dispersion (due to fabrication process) of the loop coefficients of the proposed CT $ \Delta  \Sigma $ modulator. \\
The paper is organized as follows: In Section II, we present the different arguments for the chosen architecture and the targeted specifications. In Section III, we explain the systematic high-level system method (the algorithm of the script) to analyze the the effect of component dispersion. Section IV shows the results of simulations. We finish with conclusions and perspectives in Section V.

\section{The architecture of CT $\Delta \Sigma$ MODULATOR}

\subsection{Modulator specifications}
For such application \cite{b7}, we need to design an ADC with a resolution of 10 bits and a bandwidth of 40 MHz (we decided to target slightly larger specifications in order to anticipate any cause of degradation without affecting the application). The targeted specifications of the CT $\Delta \Sigma$ ADC are resumed in table \ref{tab_01}.

\begin{table}[htbp]
\centering
\caption{Proposed specifications of the CT $\Delta \Sigma$ ADC}
\label{tab_01}
\begin{tabular}{|l|l|}
\hline
\textbf{Parameter}  & \textbf{Value}  \\ \hline
Signal bandwidth    & 40 MHz          \\ \hline
OSR                 & 8               \\ \hline
Loop filter order L & 5               \\ \hline
Modulator ADC N     & 3-Bit           \\ \hline
Sampling frequency  & 640 MHz         \\ \hline
Theoretical SQNR    & 80 dB           \\ \hline
Linearity           & \textless 0.2\% \\ \hline
\end{tabular}
\end{table}

The fact that we are designing a wide band modulator, oblige us to choose a low oversampling ratio topology in order to avoid more complexity problems in the acquisition phase; we decided to work with an OSR of eight, which means a sampling frequency of 640 MHz. To define correctly the rest of key specifications, we used equation \ref{eq_1}:
\begin{multline}
SQNR_{max}[dB]=6.02N+1.76+(2L+1)\log_{10}(OSR)+ \\
10\log_{10}(2L+1)-(2L)10\log_{10}(\pi)
\label{eq_1}
\end{multline}

It is already shown that the theoretical SQNR is usually chosen to be 10 – 20 dB higher than the required one in order to provide some tolerance for the inevitable degradation caused by the circuit non-idealities and to allow as much as possible of the noise budget for thermal noise \cite{b1,b6}.

\subsection{Architecture Selection}
\begin{figure}[htbp]
\centerline{\includegraphics[scale=0.21]{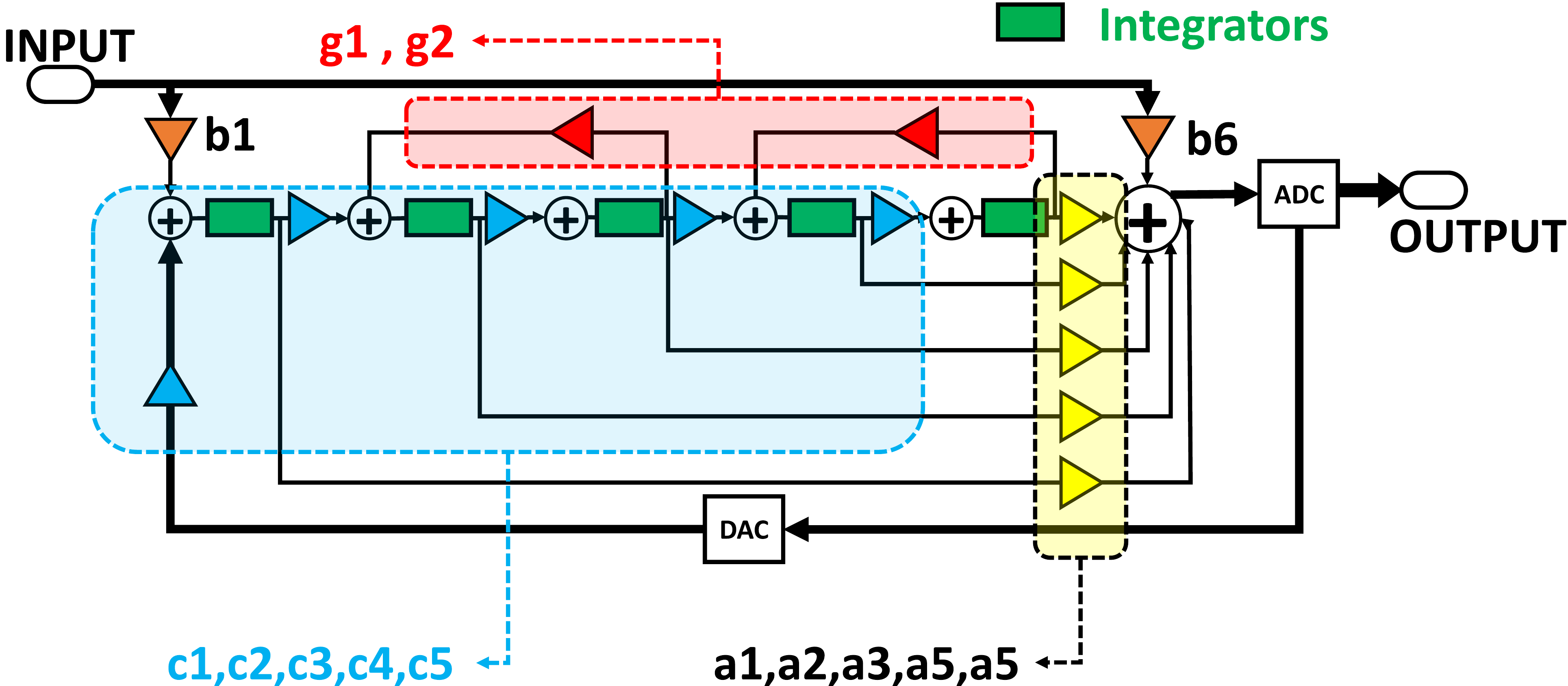}}
\caption{The CRFF Architecture of the proposed CT $ \Delta \Sigma$ modulator}
\label{fig_03}
\end{figure}

We chose to work with a Cascaded Resonators Feedforward architecture (CRFF)for the following reasons \cite{b1,b6,b8}:
\begin{itemize}
\item The first integrator of this architecture is wide band compared to other implementations which allows us to reduce both area and power consumption in terms of integrators implementations (size of capacitors).
\item The use of large bias currents in the first stage that allow us to limit noise and distortion.
\item Allow us to reach high SQNR values with low oversampling ratio (OSR) values.
\end{itemize}
Figure \ref{fig_03} illustrates a simplified block diagram of the proposed modulator. 

This architecture was implemented as a graphical model in SIMULINK as explained in \cite{b1} based on the general definition presented in \cite{b6}. It is designed to meet the targeted requirement of table \ref{tab_01}.

\section{Systematic high-level modeling}
In this section, we explain the use of the graphical behavioral simulation tool as an application of the model-based design approach. This allows to synthesize the parameters of the desired $\Delta \Sigma$  modulator using Schreier’s Toolbox. After that, we discuss the integration of a MATLAB script that control the simulation settings of the graphical model for the study of the dispersion's effect of the loop coefficients. This helps to reduce considerably the amount of time and effort applied during the schematic design phase. This method gives more information about the behavior of the topology than what is classically obtained via the use of Schreier's toolbox at the same  design phase. 

\subsection{Integration of MATLAB scripts }
\begin{figure}[htbp]
\centerline{\includegraphics[scale=0.8]{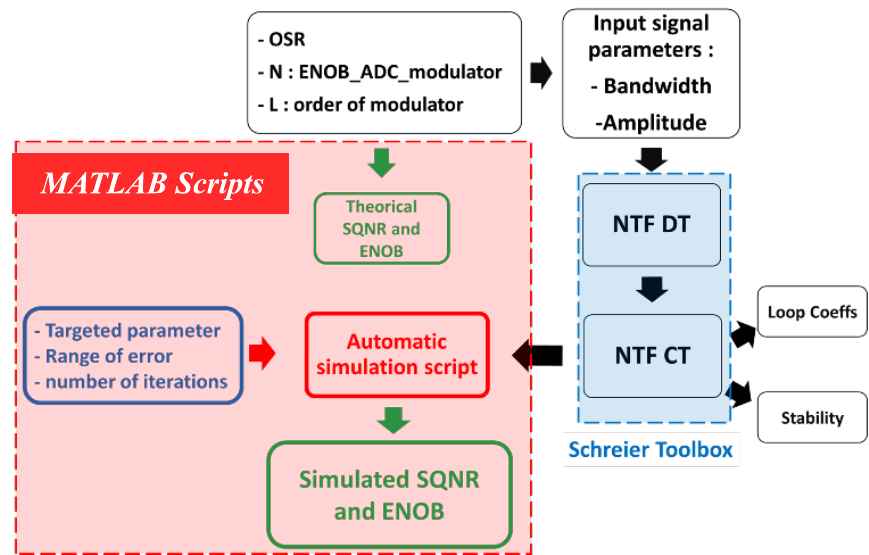}}
\caption{Diagram of our contribution to the synthesis process of CT $ \Delta \Sigma$ modulators}
\label{fig_04}
\end{figure}

The idea behind this is to optimize the use of the graphical behavioral simulation tool using MATLAB scripts as shown in figure \ref{fig_04}.  This give us access to add new features and more control over the systematic design such as the case of parametric Monte Carlo like simulations to check the effect of every parameter choice in a fast way before going to the transistor level implementation.   
Usually, this class of models are used to create a reference theoretical “ideal” model of a topology. Thus, it is always described, as “it is only useful to simply illustrate the concept of a topology.” Yet, we believe that it has more potential when we combine it with MATLAB scripts in order to estimate as soon as possible in the design flow, the impact of non-idealities such as the case of process variation. We propose the integration of scripts that control the graphical model tool focusing on two main steps:
\begin{itemize}
\item First, we evaluate the theoretical impact of the choice of the main parameters on the Signal-to-Quantization-Noise Ratio (SQNR) using equation \ref{eq_1}. In our case, this equation is calculated using MATLAB Symbolic Math Toolbox.
\item Second, we give access to the value of each parameter of every building block. Every parameter is defined as a variable in MATLAB scripts before injecting it into the graphical model. Since the model is controlled by the set of scripts, we can create any type of scenarios where parameters can vary according to a defined function with a defined number of iterations. 
\end{itemize}
\subsection{Implementation of Monte Carlo like simulations}
As an application to the feature explained above, we built Monte Carlo like simulations, where we vary independently each parameter following a normal distribution function in a chosen range of error. Using this, we can simulate in the case of this paper, the effect of component dispersion (due to fabrication process) of the loop coefficients of the proposed modulator topology.  Since we plan to use Opamp-RC integrators to meet the linearity requirement \cite{b6}, we need to estimate the time constant shift due to the process variation of R and C values. The algorithm of the script is illustrated. 

\begin{algorithm}[htpb]
\label{alg_01}
\DontPrintSemicolon
\DontPrintSemicolon
\SetAlgoLined
\SetNoFillComment
\LinesNotNumbered 
  
  \KwInput{ \\
  $a_s, b_s, c_s, g_s$ \tcp*{Calculated Coefficients using Schreier's toolbox}\\ 
  $\Delta{p},  \delta{m}$ \tcp*{Process variation error, Mismatch error} 
  $K$  \tcp*{Number of iterations, in our case $K=10000$}
    }
  \KwOutput{$a_i, b_i, c_i, g_i$ \tcp*{Coeffs adding dispersion} 
  $SQNR_i$, $ENOB_i$ \tcp*{Calculated SQNR and ENOB for every added dispersion}
  
  }
  

  \SetKwFunction{FGS}{GenerateDisp}
  \SetKwFunction{SDP}{SimulateDisp}
 
  \SetKwProg{Fn}{Function}{:}{}
  \Fn{\FGS{$a_s, b_s, c_s, g_s$, $\Delta{p}, \delta{m}$, $K$}}{
        \While{$i \leq K$}
   {
   		Pick a random value  $\Delta{p_i} \in [-\Delta{p},\Delta{p}]$ \\
   		Pick a random value  $\delta{m_i} \in [-\delta{m},\delta{m}]$\\
   		Calculate $a_i, b_i, c_i, g_i$ using equation \ref{eq_2} \\
   		Increment $i$
   		\;
   }
   
        \KwRet $a_i, b_i, c_i, g_i$\;
  }
  \;
  
  \SetKwProg{Fn}{Procedure}{:}{}
  \Fn{\SDP{$a_i, b_i, c_i, g_i$, $K$}}{
  
 	 \While{$i \leq K$}
   {
   		Inject $a_i, b_i, c_i, g_i$ in SIMULINK Model \\
   		Run Simulation using a sinus input with $f_{in}=40 MHz$\\
   		Calculate $SQNR_i, ENOB_i$ using the digital output of the model \\
   		Increment $i$
   		\;
   } 
  
   \KwRet $SQNR_i, ENOB_i$\;
  
  }
  \;

\caption{Monte Carlo like simulation for parameters dispersion}
\end{algorithm}


The variation of the coefficients, are calculated using the estimation  represented by equation \ref{eq_2}:
\begin{equation}
X_i=Xs_i \times (1+ \Delta{p_i} + \delta{m_i})
\label{eq_2}
\end{equation}
Where;
\begin{itemize}

\item $Xs_i$ is the synthesized parameter 
\item $\Delta{p_i}$ is the process variation error (around $20\%$)
\item $\delta{m_i}$ is the mismatch error (around $1\%$) 
\end{itemize}

\section{Simulations results}
In this section, we present the results of the synthesis of the modulator and the analysis of the effect of component dispersion on the final SQNR value. This study is the first step to demonstrate the use of our proposed scripts as a fast method of designing CT modulators. It allows extracting a lot of useful information as explained below. 

\subsection{Synthesis results}
\begin{figure}[htbp]
\centerline{\includegraphics[scale=0.9]{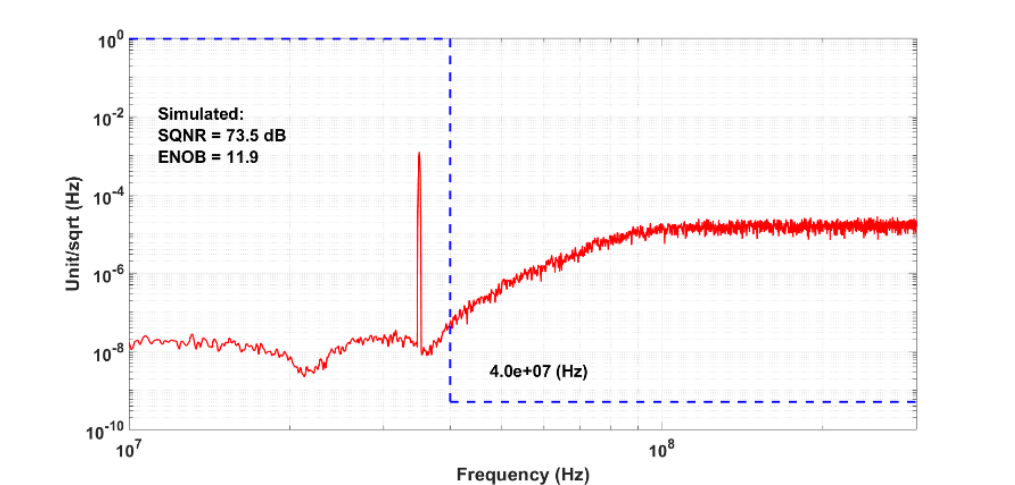}}
\caption{The SQNR of the simulated ADC }
\label{fig_05}
\end{figure}

The modulator was synthesized using the graphical behavioral simulation tool for the specifications shown in table \ref{tab_01}. Figure \ref{fig_05} shows the results. 
 \\
We achieved a simulated ENOB of $12$ bits using an input sine wave signal with a frequency of \unit{35}\mega\hertz)  and an amplitude of $0.7$ of the full input range referred to the unity as shown in figure \ref{fig_05} (As expected, we have around $-10 \hspace{0.1cm} dB$ compared to the theoretical SQNR) \cite{b1,b6,b8}. The obtained loop filter coefficients are summarized in table \ref{tab_02}. 
\begin{table}[htbp]
\caption{Coefficients of the loop filter}
\begin{center}
\label{tab_02}
\begin{tabular}{|l|l|l|l|l|}
\hline
\textbf{Coefficient} & \textbf{$a_s$} & \textbf{$g_s$} & \textbf{$b_s$} & \textbf{$c_s$} \\ \hline
1                    & 4.4392      & 0.0493      & 0.2102      & 0.2102      \\ \hline
2                    & 6.3822      & 0.2762      & 0           & 1.1023      \\ \hline
3                    & 3.4840      & -           & 0           & 0.9044      \\ \hline
4                    & 1.8533      & -           & 0           & 0.7257      \\ \hline
5                    & 0.0382      & -           & 0           & 0.4537      \\ \hline
6                    & -           & -           & 1.0000      & -           \\ \hline
\end{tabular}
\end{center}
\end{table}
These Coefficients are used to calculate values for the component used in the implementation of the modulator such as the values of resistors and capacitors. \\
Using the graphical model in SIMULINK, leads us to extract more information about what we should expect as requirement for the physical implementation. The main extracted parameters of the different Opamp-RC integrators are illustrated in table \ref{tab_03}.
\begin{table}[]
\centering
\caption{Integrators specifications}
\label{tab_03}
\begin{tabular}{|l|l|l|l|}
\hline
\textbf{Integrator} & \textbf{Maximum output swing} & DC Gain & Saturation \\ \hline
Integrator 1        & ±174 mV                       & 40   dB & 450 mV     \\ \hline
Integrator 2        & ±170 mV                       & 40   dB & 450 mV     \\ \hline
Integrator 3        & ±163 mV                       & 40   dB & 450 mV     \\ \hline
Integrator 4        & ±189 mV                       & 40   dB & 450 mV     \\ \hline
Integrator 5        & ±195 mV                       & 40   dB & 450 mV     \\ \hline
\end{tabular}
\end{table}

\subsection{Dispersion simulation}

\begin{figure}[htbp]
\centerline{\includegraphics[scale=1]{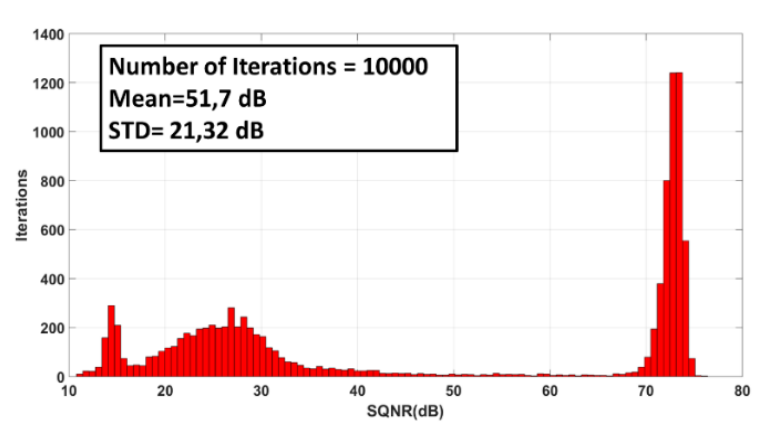}}
\caption{SQNR values for $10000$ simulations with  $\Delta{p}  =20\%$ , $\delta{m}=1\%$ on each coefficient }
\label{fig_06}
\end{figure}

In this application example, we described the variation of each coefficient of the loop individually following a normal Gaussian distribution with $\Delta{p}  =20\%$ and  $\delta{m}=1\%$. Results of $10000$ simulations are shown in figure \ref{fig_06}.

\begin{figure}[htbp]
\centerline{\includegraphics[scale=1]{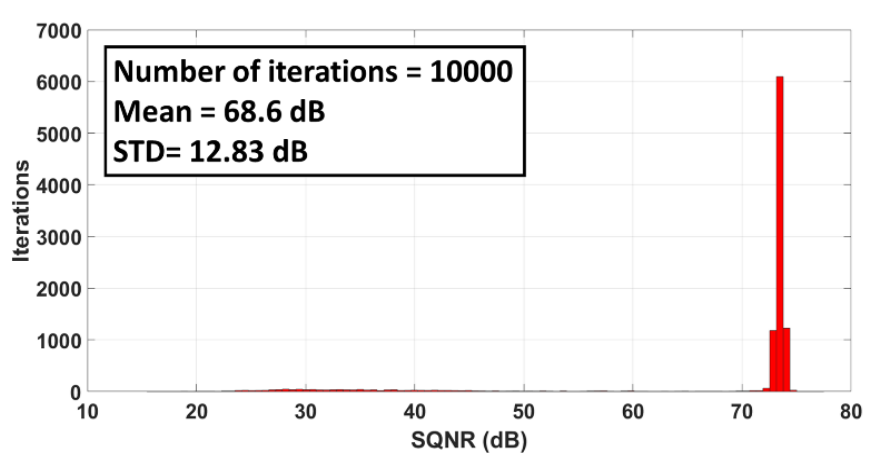}}
\caption{SQNR values for $10000$ simulations with  $\Delta{p}  =10\%$ , $\delta{m}=0.5\%$ on each coefficient}
\label{fig_08}
\end{figure}

We can see that the modulator is not immune to this level of dispersion, which leads to many questions. 
In order to investigate this, we did a series of simulations to find the most critical parameters that cause this behavior. This time, we applied the same dispersion on one parameter a time, and for every simulation, and we found that the coefficients $a_1, a_2, b_1, c_1$, which are the most critical ones. We doubted that the main reason of this drop of SQNR is the saturation of integrators. 
To validate this hypothesis, we only reduced the range of $\Delta{p}  =10\%$ , $\delta{m}=0.5\%$  for these critical parameters. Results of this simulation are presented in figure \ref{fig_08}. 

We concluded that in order to guarantee the desired specifications, these parameters must be correctly designed within a range of error as low as possible.

\section{Concusion}

In this paper, we discussed the development of a systematic high-level design a fifth order CT $\Delta \Sigma $ modulator for 10-bit ENOB ADC based on a CRFF architecture. The following results are reported here:
\begin{itemize}
\item We showed the use of the graphical behavioral models in the flow of design of CT $\Delta \Sigma$ modulators.
\item We explained the synthesis of a fifth order CT $\Delta \Sigma$ modulator for 10-bit ENOB ADC based on a CRFF architecture.
\item We presented the results of the study of the effect of the dispersion (due to fabrication process) of the loop coefficients taking into consideration the non-idealities of Opamp-RC integrators.
\end{itemize}
More non-idealities of the other blocks will be added to the model (DAC non-linearity, Jitter...etc.) which allow us to estimate more precisely the SQNR and calculate more key performances as the SNDR. This extracted data is used to design the building blocks using fast design methodologies such as:$\frac{g_m}{I_d}$  \cite{b9}.

\section*{Acknowledgment}

We thank DIAMASIC project for financial and technical supports.

\end{document}